\documentclass{an_2c}
\input psfig.tex

\begin{document}
\sloppy

\title{X-ray Rapid Variability of MKN 421}

\author{A.Treves\inst{1}, Y.H. Zhang\inst{2}, F.Tavecchio\inst{3}, 
A. Celotti\inst{2}, 
L. Chiappetti\inst{4}, 
G. Fossati\inst{5}, G. Ghisellini\inst{3}, L. Maraschi\inst{3}, E. Pian\inst{6}, 
G. Tagliaferri\inst{3}.}
\institute{Universita' dell'Insubria, Como, Italy
\and 
SISSA/ISAS, Trieste, Italy
\and 
Osservatorio Astronomico di Brera, Milano-Merate, Italy
\and
IFC/CNR, Milano, Italy
\and
CASS/UCSD, La Jolla, USA
\and
ITeSRE/CNR, Bologna, Italy}
\headnote{Astron. Nachr. 320 (1999)}
\maketitle


The bright close-by BL Lac object MKN 421 was observed at several epochs
with BeppoSAX. Here we concentrate on the campaign of April-May 1997
(Fossati et al 1998, Guainazzi et al 1999), and on that of April 1998, 
which was simultaneous with a TeV flare (Maraschi et al 1999a, b, Fossati et 
al 1999).
 
We consider rapid variability focussing on the issue of lags between  
a soft band (LE: 0.1-1.5 keV) and a medium energy band (ME: 3.5-10 keV). 
Light curves are reported in the quoted references. The analysis follows the 
procedures adopted for PKS 2155-304 (Zhang et al 1999). Lags have been 
searched for using the Discrete Correlation Function (DCF), and the Modified Mean 
Deviation (MMD) techniques (e.g. Edelson et al 1995). The resulting distributions 
are fitted with Gaussians. The position of the peak is taken as the lag, 
positive if ME precedes LE. Following Peterson et al (1998), the 
uncertainties depending on photon statistics are evaluated through 
Monte Carlo simulations, considering "flux randomization" and "random subset 
selection", see Fig. 1. From the distribution the 90\% confidence intervals 
are deduced. Results are summarized in Table 1. The discovery of a negative
lag in 1998 observation and the evaluation of its uncertainty appears as
a noticeable result.
 
In the quoted observations of PKS 2155-304 
the lag was always positive and its value decreased with increasing source 
intensity. 
The 1998 state of MKN 421 ($F_{2-10 ~ keV}\simeq 3\times 10^{-10}$ 
erg cm$^{-2}$ s$^{-1}$) was indeed quite higher than that of 1997
($F_{2-10 ~ keV} \simeq 0.7\times 10^{-10}$ erg cm$^{-2}$ s$^{-1}$). The 
sign inversion may be related to the same phenomenology noted in PKS2155-304 
(see the GINGA observation reported in Sembay et al. 1993). 
An interpretative picture requires a rather complex model on the line
of those proposed by Tashiro et al. (1995), Dermer (1998), Kirk et al. (1998), 
Chiaberge \& Ghisellini (1999), Georganopoulos 
\& Marscher (1999) and Kataoka et al. (1999).

\begin{figure}
\centerline{\psfig{figure=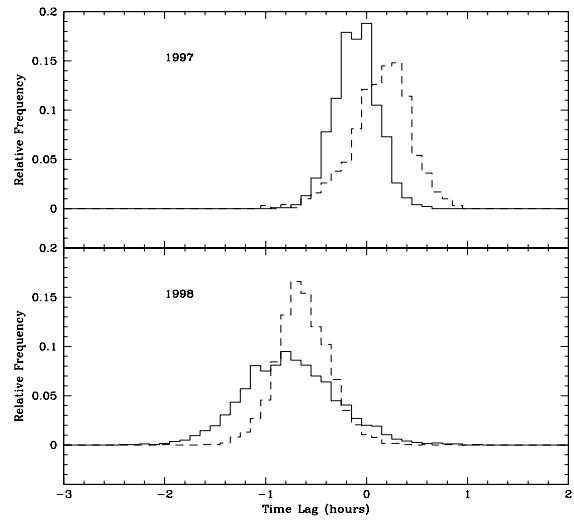,width=8 truecm,height=7.5 truecm,clip=}}
\caption{Results of the Monte Carlo simulations for the DCF method (solid line)
and the MMD method (dashed line) for the 1997 and 1998 data.}
\end{figure}

\begin{table}
\begin{center}
\begin{tabular}{cccccc}
\hline \hline
&& \multicolumn{2}{c} {DCF} & \multicolumn{2}{c} {MMD}\\ 
&& Lag & 90\% C.L. & Lag & 90\% C.L.\\ \hline 
&&&&&\\ 
\multicolumn{6}{c} {\bf Mkn 421 $Beppo$SAX 1997} \\ \hline
Gaussian fit & & -0.19& -0.26,-0.11 
                      & 0.16& 0.04,0.27\\
MC Simulation & & -0.10& -0.45,0.23
                      & 0.14& -0.39,0.59\\
&&&&&\\
\multicolumn{6}{c}{\bf Mkn 421 $Beppo$SAX 1998} \\ \hline
Gaussian fit && -0.82& -0.91,-0.73 
		       & -0.60 & -0.73,-0.47\\
MC Simulation && -0.75 & -1.48,0.02 
		      & -0.62 & -1.06, -0.17\\
&&&&&\\
\hline
\end{tabular}
\caption{Results from Cross Correlations and Monte Carlo simulations
for the 1997 and 1998 $Beppo$SAX observations of Mkn 421. The lags (in hours)
and the 90\% confidence levels are given for both methods.}
\end{center}
\end{table}

\end{document}